\documentclass[conference]{IEEEtran}
\IEEEoverridecommandlockouts

\usepackage{cite}
\usepackage{url}
\usepackage{amsmath,amssymb,amsfonts}
\usepackage{algorithm}
\usepackage{algpseudocode}
\usepackage{graphicx}
\usepackage{textcomp}
\usepackage{xcolor}
\usepackage{xspace}
\usepackage{comment}
\usepackage{kotex}
\usepackage[normalem]{ulem}
\def\BibTeX{{\rm B\kern-.05em{\sc i\kern-.025em b}\kern-.08em
    T\kern-.1667em\lower.7ex\hbox{E}\kern-.125emX}}
\begin{document}

\newcommand{\ArchName}{ABC-FHE\xspace}
\newcommand{\CoreName}{RSC\xspace}
\title{\LARGE \textbf{\ArchName : A Resource-Efficient Accelerator Enabling Bootstrappable Parameters for Client-Side Fully Homomorphic Encryption\\[-1em]}
}

\author{
    \IEEEauthorblockN{Sungwoong Yune \quad Hyojeong Lee \quad Adiwena Putra \quad Hyunjun Cho \\
    \quad Cuong Duong Manh \quad Jaeho Jeon \quad Joo-Young Kim}
    \IEEEauthorblockA{
        School of Electrical Engineering, KAIST\\
        \{imwooong, hjlee8877, adiwena.research, h.cho, cuongdm1410, math15738, jooyoung1203\}@kaist.ac.kr
    }\\[-3em]
}





\maketitle

\begin{abstract}

As the demand for privacy-preserving computation continues to grow, fully homomorphic encryption (FHE)—which enables continuous computation on encrypted data—has become a critical solution. However, its adoption is hindered by significant computational overhead, requiring 10000-fold more computation compared to plaintext processing. Recent advancements in FHE accelerators have successfully improved server-side performance, but client-side computations remain a bottleneck, particularly under bootstrappable parameter configurations, which involve combinations of encoding, encrypt, decoding, and decrypt for large-sized parameters. To address this challenge, we propose \ArchName, an area- and power-efficient FHE accelerator that supports bootstrappable parameters on the client side. \ArchName employs a streaming architecture to maximize performance density, minimize area usage, and reduce off-chip memory access. Key innovations include a reconfigurable Fourier engine capable of switching between NTT and FFT modes. Additionally, an on-chip pseudo-random number generator and a unified on-the-fly twiddle factor generator significantly reduce memory demands, while optimized task scheduling enhances the CKKS client-side processing, achieving reduced latency. Overall, \ArchName occupies a die area of 28.638 mm$^2$ and consumes 5.654 W of power in 28 nm technology. It delivers significant performance improvements, achieving a 1112$\times$ speed-up in encoding and encryption execution time compared to a CPU, and 214$\times$ over the state-of-the-art client-side accelerator. For decoding and decryption, it achieves a 963$\times$ speed-up over the CPU and 82$\times$ over the state-of-the-art accelerator.

\end{abstract}


\vspace{-0.1in}

\section{Introduction}
\label{section1}

In recent years, personal data from edge devices, such as PCs and mobile phones, has been increasingly sent to centralized servers for advanced computation with the paradigm of cloud computing~\cite{satyanarayanan2017emergence}. This trend is evident in real-time AI applications that have excessive computational requirements, including chatbot~\cite{chen2017survey}, voice assistant~\cite{kepuska2018next}, and image recognition~\cite{krizhevsky2012imagenet}. However, this raises privacy concerns, as users should share their personal data on cloud servers. Fully homomorphic encryption (FHE) is considered the solution by allowing computations on data while it remains encrypted, thereby preserving privacy.

While FHE provides a solution for privacy-preserving computations, it also introduces substantial computational challenges~\cite{lee2022privacy}. Specifically, in the CKKS~\cite{cheon2017homomorphic} scheme, a popular choice for AI-based applications, the ciphertext is a high-degree polynomial (ranging from degree $2^{14}$ to $2^{16}$) with each coefficient represented as a high-precision integer of hundreds to thousands of bits. These coefficients are decomposed in the residue number system (RNS), reducing them to approximately 50–60 bits and enabling more efficient arithmetic operations. To accelerate these computations, the number theoretic transform (NTT) is often used, reducing the complexity of polynomial multiplication from $N^{2}$ to $N log(N)$. Despite this algorithmic enhancement, workloads within the FHE domain remain up to 10,000$\times$ slower than their plaintext counterparts due to the inherent computational complexity, emphasizing the need for dedicated hardware accelerators to achieve the demanding performance of modern applications~\cite{jung2021accelerating}.
\begin{figure}[t]
    \centering
    \includegraphics[width=3.4in]{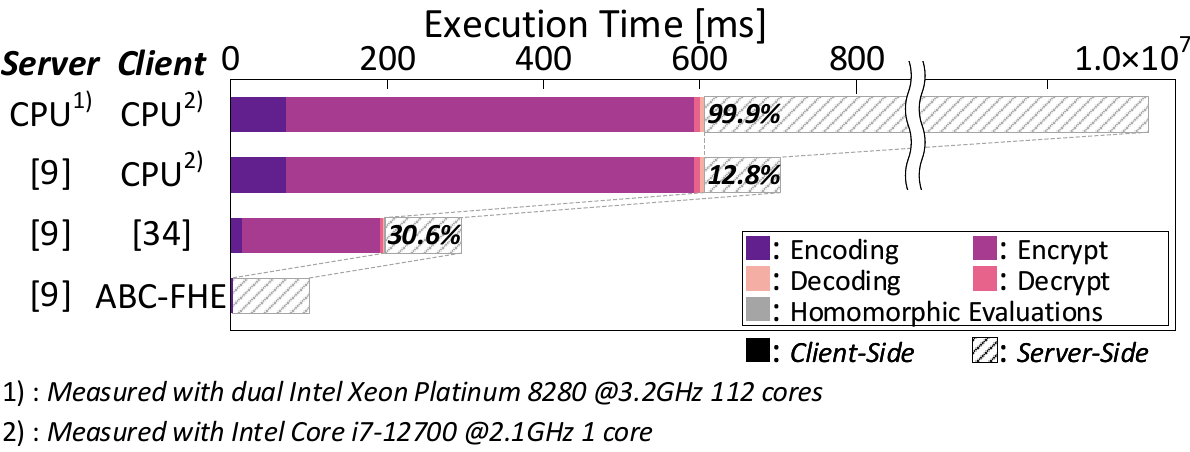}
    \vspace{-0.10in}
    \caption{Execution time breakdown of overall client-side and server-side operations.}
    \label{fig:introduction}
    \vspace{-0.3in}
\end{figure}





Among hardware accelerators, the majority have focused on achieving advancements on the server-side~\cite{kim2022ark, kim2023sharp, putra2023strix, putra2024morphling, deng2024trinity}. However, the need for client-side accelerators has become increasingly evident for two main reasons. First, server-side computation is no longer the primary bottleneck due to drastically reduced computation times by server-side FHE accelerators. For example, as shown in Figure \ref{fig:introduction}, executing FHE operations on the ResNet20 model reveals that encryption and decryption tasks on the client-side account for 69.4\% of the total execution time using a state-of-the-art (SOTA) client-side accelerator~\cite{wang2024compact}, whereas computations on the server-side account for only 30.6\% with a SOTA server-side ASIC accelerator~\cite{deng2024trinity}. Second, client-side computations involve distinct workload characteristics, requiring both fast Fourier transform (FFT) and NTT for processing. These computations depend on floating-point and integer operations, respectively, and are typically performed in a non-repetitive manner, presenting unique challenges compared to server-side workloads.

Several prior works have explored client-side FHE accelerators~\cite{azad2022race, di2023vlsi, krieger2024aloha, wang2024compact}, but they suffer from several limitations. First, these accelerators are constrained to small FHE parameters (e.g., $N=2^{13}$), rendering them incapable of supporting bootstrapping operations that require larger polynomial degrees, typically at least $N=2^{14}$ and often extending to $N=2^{16}$. Second, prior works using non-streaming architectures encounter severe DRAM bandwidth limitations. Even if throughput is increased to accelerate computations, the resulting larger data output per cycle exceeds the available DRAM bandwidth, leading to latency overhead rather than performance gains. Third, the SOTA work~\cite{wang2024compact} intensifies the DRAM bandwidth bottleneck by directly fetching parameters from DRAM, further increasing memory bandwidth demands and exacerbating latency overhead.

To address these issues, we propose \ArchName, a resource-efficient FHE accelerator specifically designed to support bootstrappable parameters for client-side applications. First, it is tailored to accommodate bootstrappable parameters, aligning with the practical requirements of FHE deployment. Second, it adopts a streaming architecture, addressing memory bottlenecks associated with large bootstrappable parameters with efficient memory utilization and seamless operation. Third, it incorporates an on-chip pseudo-random number generator (PRNG) and an on-the-fly twiddle factor generator (OTF TF Gen) to generate random values and twiddle factors directly on-chip, reducing reliance on external memory. Furthermore, it introduces methods to minimize the area overhead caused by the structural characteristics of the streaming architecture.
To summarize, we make the following contributions:
\begin{itemize}
    \item We analyze the workload of the client-side CKKS scheme, identifying distinct workload characteristics and imbalances between encryption-related tasks (IFFT + NTT) and decryption-related tasks (FFT + INTT).
    \item To meet latency requirements while accommodating diverse workload characteristics, we propose a streaming architecture with a reconfigurable compute engine that supports both floating-point complex-number-based I/FFT and integer-based I/NTT operations.
    \item  To further optimize area efficiency, we introduce the unified on-the-fly twiddle factor generator (OTF TF Gen), enabling the NTT/FFT engine to share a single hardware unit for twiddle factor generation. Additionally, we provide a methodology for selecting optimal prime numbers to facilitate hardware-friendly modular multiplication.
    \item Overall, \ArchName achieves a speed-up of up to 1112× for encoding and encryption and up to 963× for decoding and decryption compared to a CPU, while delivering up to 214× and 82× improvements, respectively, over SOTA accelerators.
\end{itemize}
\vspace{-0.1in}

\section{background}
\label{section2}

For the remainder of this paper, we use the following notations: 
In representation of polynomial, \(n\) denotes the degree, \(N\) represents the polynomial degree, \(Q\) refers to the prime number, and \(P\) in the FFT architecture indicates the number of parallelism lanes.

\subsection{Fully Homomorphic Encryption}
FHE achieves security by injecting error into the computation process, requiring a specialized operation called bootstrapping to reset the accumulated error during consecutive computations. In the CKKS scheme, this error is managed as ``levels" which decrease as error accumulates through operations. Bootstrapping restores the levels and reduces the error; however, the process also consumes levels as part of its execution. To minimize the frequency of this computationally intensive bootstrapping, higher levels are necessary. Achieving a high level while ensuring the 128-bit security standard, which is considered secure in modern cryptographic systems, requires large polynomial degrees ranging from $2^{14}$ to $2^{16}$~\cite{bossuat2024security}.

\subsection{Fourier-like Transform}
In FHE, polynomial multiplication is required, which has a complexity of $N^{2}$. By transforming from the time domain to the frequency domain using fast transforms, this complexity can be reduced to $N \log(N)$. Two types of transformations are commonly used in FHE: NTT and FFT, which perform complex-number and modular computations, respectively. Both transforms share the same fundamental equation but differ in their respective weights, known as the twiddle factors. Specifically, NTT operates within a prime modular ring (Q), while FFT operates on the unit circle in the complex plane. The equation for Fourier-like transform is as follows :
\[
A[k] = \sum_{i=0}^{N-1} a(i) \times W_N^{ki} \quad \cdots (1)
\]

where \( W_N \) represents the twiddle factor for each transformation.

\begin{figure}[t]
    \centering
    \includegraphics[width=3.4in]{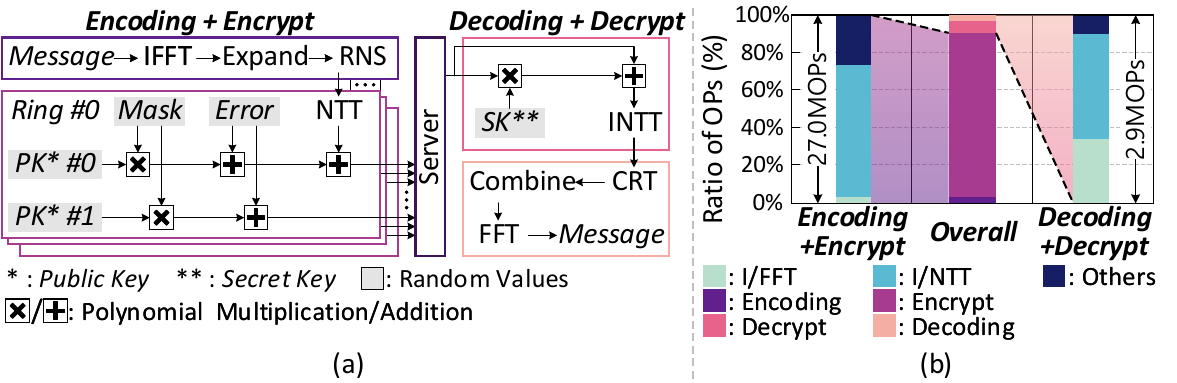}
    \vspace{-0.10in}
    \caption{Workload analysis of CKKS client-side operations: (a) Operational flow and (b) Ratio of operations for 12 level encoding/encryption and 1 level decoding/decryption supporting polynomial degree $2^{16}$.}
    \label{fig:Background}
    \vspace{-0.25in}
\end{figure}




\begin{figure*}[t]
    \centering

    \includegraphics[width=7.0in]{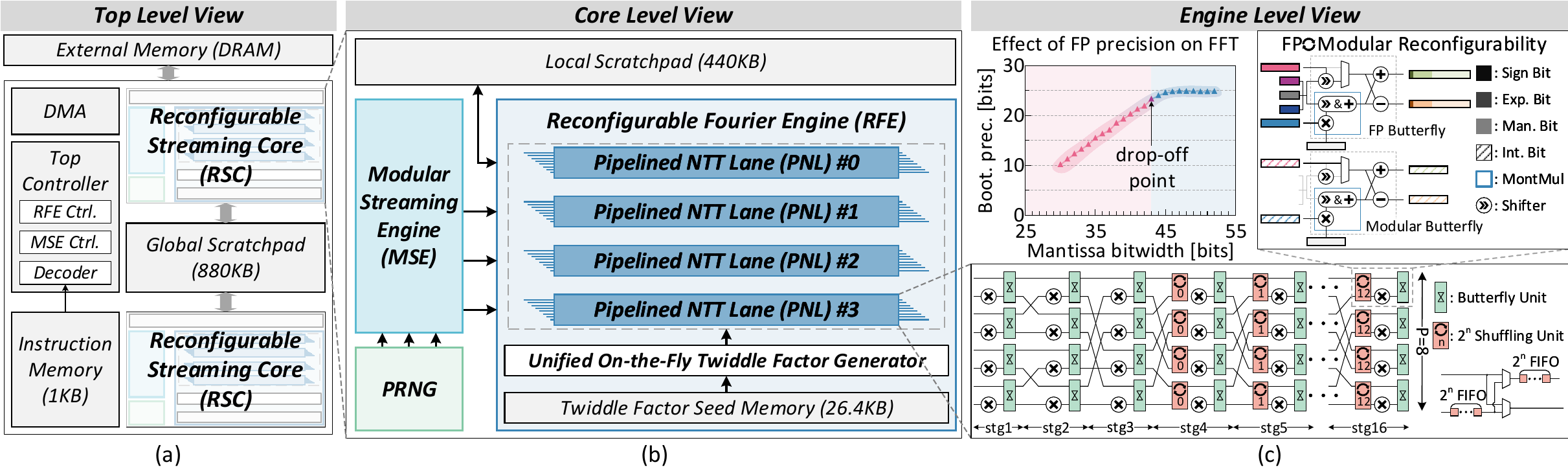}
    \vspace{-0.10in}
    \caption{Overall architecture of \ArchName (a) Top level view (b) Core level view (c) Engine level view.}
    \label{fig:Overall_Arch}
    \vspace{-0.25in}
\end{figure*}




\subsection{Hardware for Fourier-like Transform}
Even with complexity optimization through FFT, $N$ can reach up to $2^{16}$ in CKKS, which still requires high computational capacity. Many prior works have sought to support the FFT algorithm through dedicated hardware~\cite{johnston1983parallel, ahmed2013low, garrido2016serial}. There are two main approaches to optimizing FFT architecture.

    \textbf{Spatial Approach.}
    To maximize throughput, increasing the number of computation lanes (P) is an obvious approach, as it accelerates processing at the cost of additional area. However, on the client-side, where careful external memory accesses (EMA) handling is required, generating a large amount of output data per cycle can lead to additional stalls. Therefore, the number of lanes must be chosen carefully, as a larger number is not always optimal.
    
    \textbf{Temporal Approach.}
    The limitation on the number of lanes necessitates an architecture with multiple stages. Each stage operates in a pipelined manner, allowing the Fourier-like transform to achieve constant throughput per cycle, eliminating any latency disparities compared to non-pipelined designs. This makes a pipelined NTT architecture more suitable for client-side hardware.

\subsection{Workload Characteristic of Client-Side FHE}
Figure \ref{fig:Background} highlights two key characteristics of CKKS client-side operations when computing a single ciphertext. First, as shown in Figure \ref{fig:Background} (a), encoding and encryption require simultaneous support for IFFT and NTT to convert messages into ciphertexts, whereas decoding and decryption require FFT and INTT for reversing ciphertexts into messages. Second, as illustrated in Figure \ref{fig:Background} (b), the number of operations for encoding and encryption is nearly ten times greater than for decoding and decryption. This imbalance underscores the inefficiency of implementing separate modules for encryption and decryption.

\section{\ArchName Architecture}
\label{section3}

To address the limitations of prior works that are restricted to small parameters~\cite{azad2022race, di2023vlsi, krieger2024aloha, wang2024compact}, we designed an accelerator capable of efficiently handling larger parameters. In resource-constrained client-side environments, simply increasing throughput without regard for area and resource limitations can lead to memory bottlenecks or an unrealistic demand for on-chip memory. Thus, \ArchName employs a streaming architecture to maintain consistent throughput while optimizing area, addressing these challenges effectively.

Figure \ref{fig:Overall_Arch} (a) shows the overall architecture of \ArchName. At the top level, two homogeneous reconfigurable streaming cores (\CoreName) enable the streaming-based design to stably handle either message encoding or ciphertext decoding. The reconfigurable nature of \CoreName allows for three operational modes: doubling the throughput for encrypt, doubling the throughput for decrypt, or simultaneously performing encrypt and decrypt. Additionally, a global scratchpad is included to fetch messages or ciphertext from external memory, ensuring continuous data flow to the two cores.

Figure \ref{fig:Overall_Arch} (b) presents the core-level architecture of \CoreName, optimized for efficient streaming operations across its three operational modes. At its core, \CoreName incorporates two primary engines to enable seamless functionality: the reconfigurable Fourier engine (RFE) and the modular streaming engine (MSE). The RFE, equipped with four pipelined NTT lanes (PNLs), supports both I/NTT and I/FFT computations to maximize throughput for Fourier-like transforms. Complementing this, the MSE handles additional SIMD-like operations, including RNS, Chinese Remainder Theorem (CRT) computations for inverse RNS, and element-wise additions and multiplications. To minimize EMA, the architecture emphasizes on-chip value generation. Specifically, an on-chip PRNG produces random values such as masks, errors, and keys for ciphertext generation, eleminating the need to fetch them from external memory. While the OTF TF Gen, shared among the four PNLs, dynamically generates twiddle factors for Fourier-like transforms.

Figure \ref{fig:Overall_Arch} (c) highlights the RFE, the central component of \CoreName, designed to efficiently support both I/FFT and I/NTT operations through optimized bitwidth configurations. For NTT, recent studies addressed the challenge of substantial bitwidth requirements in CKKS by dividing data into smaller 32–36 bit segments across multiple levels while doubling the number of levels, a solution proven feasible for CKKS~\cite{agrawal2023high}. This allows \ArchName to adopt a smaller datapath bitwidth, optimizing efficiency. For I/FFT, required precision is measured by bootstrapping precision (Boot. prec.), which indicates the bit precision needed after server-side bootstrapping to maintain model accuracy. Prior research~\cite{kim2023sharp} established that a Boot. prec. above 19.29 bits sufficiently supports AI models without significant accuracy degradation. To minimize the FFT datapath, which traditionally relies on the 64-bit floating-point (FP64) data type, we iteratively reduced the floating-point mantissa bitwidth and evaluated Boot. prec. at each step. As shown in the graph in Figure \ref{fig:Overall_Arch} (c), maintaining at least 43 mantissa bits results in a Boot. prec. of 23.39 bits, exceeding the required threshold. Accordingly, we implemented a custom 55-bit floating-point (FP55) format for I/FFT and a 44-bit modular operation for I/NTT, designed in a reconfigurable manner to maximize efficiency for both computation types.

\section{Microarchitecture}
\label{section4}
%



\subsection{Reconfigurable Fourier Engine (RFE)}
\ArchName employs the NTT algorithm to perform polynomial multiplication. To achieve acceleration through increased throughput while maintaining a consistent rate, a pipelined NTT design with an multi-path delay commutator (MDC)-based backbone~\cite{johnston1983parallel} (P=8) was chosen. However, the increased number of butterfly units required for this design results in larger overall area. To address this, \ArchName introduces three optimizations: (1) twiddle factor scheduling to reduce the total number of multipliers, (2) an algorithm-hardware co-designed Montgomery multiplier to minimize multiplier area, and (3) leveraging reconfigurability among PNLs to support both NTT and FFT operations.

    \textbf{Twiddle Factor Scheduling.}
    \begin{figure}[t]
    \centering
    \includegraphics[width=3.4in]{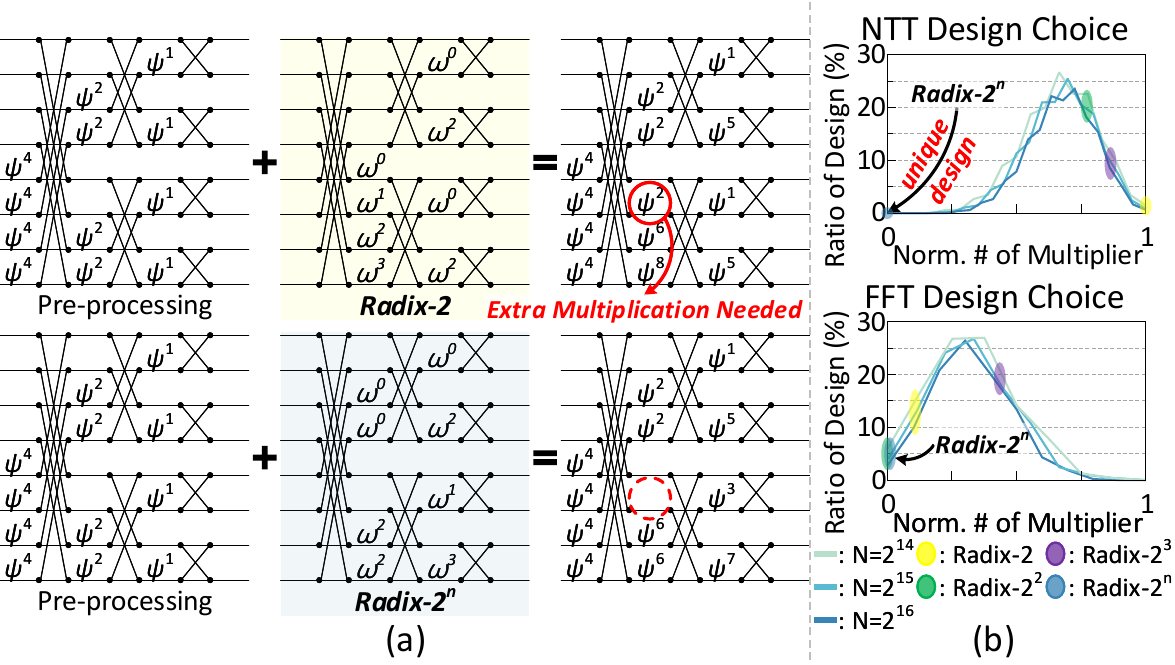}
    \vspace{-0.10in}
    \caption{(a) Twiddle factor scheduling in NTT signal flow graph (b) Distribution of multiplier counts across possible NTT/FFT design configurations.}
    \label{fig:tf_scheduling}
    \vspace{-0.25in}
\end{figure}
    Research on pipelined architectures for the FFT has been extensive~\cite{garrido2017feedforward, jang2018area, garrido2020optimum}, with a particular focus on minimizing the area of computational units. In FFT, efforts to optimize area include treating the multiplication between twiddle factors and inputs through rotators, enabling the identification of optimal rotator conditions in pipelined architectures. However, in the NTT, all multipliers are unified as modular multipliers, unlike the FFT approach. This makes it essential to reduce the number of computational units. Additionally, to simplify polynomial operations in NTT, the nega-cyclic property is employed, which requires extra pre-processing and post-processing steps. The equations for pre-processing and post-processing are shown below.
    \[
    \begin{aligned}
    A[k] & = \sum_{n=0}^{N-1} a(n) \times W_N^{kn} \times \psi_{2N}^{n} & \pmod{Q} \cdots (2) \\
    a[n] & = N^{-1} \sum_{k=0}^{N-1} A[k] \times W_N^{-kn} \times \psi_{2N}^{-k} & \pmod{Q} \cdots (3)
    \end{aligned}
    \]
    \[
    \text{where} \quad \psi_{2N}^2 \equiv W_N \pmod{Q}
    \] 
    The NTT algorithm can be optimized by merging pre-processing and post-processing operations with twiddle factors, as demonstrated in \cite{roy2014compact} and \cite{poppelmann2015high}. This merging technique eliminates the need for additional multipliers by applying a consistent pattern of twiddle factor operations across stages in the signal flow graph. By aligning the arrangement of twiddle factors with the pre-/post-processing pattern, the merging technique can be seamlessly implemented without extra computational units. Consequently, the minimum number of multipliers required for a pipelined NTT architecture is theoretically \(P/2 \times \log_2N\). Figure \ref{fig:tf_scheduling} (a) shows the distribution of twiddle factors in the signal flow graph using an 8-point example. For instance, with a radix-2 design, pre-processing results in 13 twiddle factor multiplications, while a radix-\(2^n\) design reduces this to 12 multiplications. As the polynomial degree increases, the design choice significantly impacts the total number of multiplications required.
    
    To further optimize multiplier usage, we analyzed a range of design configurations, including less conventional options, to evaluate their overheads. Figure \ref{fig:tf_scheduling} (b) compares the overheads of different design options for NTT and FFT. Our analysis revealed that only radix-\(2^n\) designs maintain the consistent twiddle factor pattern required for efficient pipelined NTT implementation, making them the most suitable choice. Consequently, we adopted radix-\(2^n\) designs for both NTT and FFT operations, highlighting that our radix-\(2^n\) configuration achieves a 29.7\% and 22.3\% reduction in number of multipliers compared to widely used radix-2 and radix-\(2^2\) designs in NTT, respectively.

    \textbf{NTT-Friendly Montgomery Multiplier.}
    Modular multiplication, which involves both multiplication and division by a specific number, is inherently expensive. To address this, both algorithmic optimizations~\cite{barrett1986implementing, montgomery1985modular} and hardware implementations have been explored~\cite{huang2008optimized, dai2016area, kim2019fpga, nguyen2023ckks}. Among the common approaches, Barrett multiplication approximates division with a multiplication to simplify the operation, while Montgomery multiplication first transforms the input into the Montgomery domain via pre-processing, performs modular reduction using simpler arithmetic, and then converts it back to the original domain via post-processing.
    Montgomery reduction efficiently reduces modular arithmetic computations in the Montgomery domain. It operates on two integers \( a \) and \( b \) (\( a, b < Q \)), with a modulus \( Q \) of \( p\_bw \)-bit bitwidth and a coprime \( R \) (\( R \geq Q \)).
    \vspace{-0.08in}
    \[
    \begin{aligned}
        T & = a \times b \quad && \cdots (4) \\
        m & \equiv ((T \pmod{R}) \times QInv) & \pmod{R} \quad & \cdots (5) \\
        t & \equiv (T - m \times Q) & \pmod{R} \quad & \cdots (6) \\
        t & = 
        \begin{cases} 
        t + Q, & \text{if } t < 0 \\
        t, & \text{otherwise}
        \end{cases} \quad && \cdots (7)
    \end{aligned}
    \]
    \vspace{-0.08in}
    
    Montgomery multiplication is typically the simplest and most efficient. However, Montgomery reduction inherently requires three multipliers, introducing significant overhead, particularly in FHE applications with large bitwidths. To address this issue, we optimized the Montgomery reduction algorithm by modifying the computation of $QInv$ to be processed using a shift-and-add approach. 

    Before presenting the optimized algorithm, we define the NTT-friendly prime number and Euler's theorem as follows:
    \[
    \begin{aligned}
        Q & = 2^{p\_bw} + k \cdot 2^{n+1} + 1 \quad & \cdots (8) \\
        QInv & \equiv Q^{2^{n-1}-1} \pmod{2^n} \quad & \cdots (9)
    \end{aligned}
    \]
    
    Equation (8) represents the expression of an NTT-friendly prime, where \( k \) is an arbitrary value. Equation (9) uses Euler's theorem to define \( QInv \). \( \equiv \) in equation (10) and (11) represent modular equivalence with \( 2^n \). Substituting the expression for \( Q \) from Equation (8) into Equation (9) yields the following:
    \[
    QInv \equiv \sum_{i=0}^{2^{n-1}-1} \binom{2^{n-1}-1}{i} \cdot (2^{p\_bw} + k \cdot 2^{n+1})^i \quad \cdots (10)
    \]

    If \( k \) satisfies the condition \( k \geq 2^{{bw/2} - 1 - n} \), with \( k = \pm 2^a \pm 2^b \pm 2^c \), \( QInv \) can be simplified as follows:
    \[
    QInv \equiv -2^{p\_bw}-(\pm2^a\pm2^b\pm2^c)\cdot2^{n+1}+1 \quad \cdots (11)
    \]

    By applying this to the Montgomery algorithm, all multiplications except for the initial multiplication can be converted into shift-and-add operations. This significantly reduces the hardware complexity compared to the original implementation, which required three multipliers.
    
    \begin{table}[t]
\centering
\caption{Area of Modular Multiplier}
\label{table:Modular_Mult_Area}
\begin{tabular}{lcc}
\hline
\textbf{Algorithm} & \textbf{Area ($\mu$m$^2$)} & \textbf{Pipeline Stages (Cycles)} \\ \hline
Vanilla Barrett   & 35054   & 4    \\
Vanilla Montgomery & 19255   & 3    \\
NTT-Friendly Montgomery & 11328  & 3    \\ \hline
\vspace{-0.35in}
\end{tabular}
\end{table}

    Despite reducing the number of supported primes due to this optimization, the design still sufficiently supports 20-40 encryption levels. For instance, in the case of $N = 2^{16}$, the required 32-36 bit primes amount to a total of 443, which is more than adequate to handle all practical scenarios. Table \ref{table:Modular_Mult_Area} further highlights the area and pipeline stages of different modular multipliers (MM), synthesized at 600 MHz using 28 nm process. The NTT-friendly Montgomery MM achieves a 67.7\% area reduction compared to the Barrett MM and a 41.2\% reduction compared to the vanilla Montgomery MM.
    
    \textbf{Reconfigurablility among PNLs.}
    To support FFT computations, complex number FP operations are required. These involve calculations such as:
    \[
    (a + bi) \cdot (c + di) = (ac - bd) + i(ad + bc) \quad \cdots (12)
    \]

    which require four FP multiplications. To efficiently handle these operations, the modular multipliers and FP multipliers are designed to operate in a reconfigurable manner. This enables the use of four modular multipliers to support complex number FP multiplication. The same approach is applied to the unified OTF TF Gen, ensuring seamless support for generating twiddle factors.

\subsection{Efficient Memory Handling}
\ArchName addresses the challenge of minimizing memory requirements for FHE operations on the client-side by generating essential data on-chip, thereby reducing reliance on external memory. For client-side FHE with a polynomial degree of $2^{16}$, 44-bit precision, and 24 levels, calculations estimate the need for 16.5 MB of public key storage, 8.25 MB for masks and errors, and an additional 8.25 MB for twiddle factors. Such substantial on-chip memory is impractical for client-side applications, while frequent external memory fetches require high-bandwidth solutions, such as HBM, which are unsuitable for typical client-side environments.

To address these constraints, \ArchName incorporates a PRNG and an unified OTF TF Gen to generate data directly on-chip. The PRNG requires only a 128-bit seed on-chip to meet the 128-bit security level, eliminating the need to store extensive precomputed values. The unified OTF TF Gen dynamically generates twiddle factors for each stage of the pipelined NTT lane using a compact twiddle factor seed and step size, requiring approximately 27 KB, which reduces on-chip memory requirements by over 99.9\%.
This efficient approach not only significantly reduces memory overhead but also enables scalable and practical FHE acceleration for client-side environments.
\section{Evalutation}
\label{section5}

\subsection{Hardware Modeling}
\begin{table}[t]
\centering
\caption{The area and power breakdown of \ArchName}
\label{table:area_power_breakdown}
\begin{tabular}{lcc}
\hline
\textbf{Component} & \textbf{Area (mm$^2$)} & \textbf{Power (W)} \\ \hline
4× PNL & 10.717 & 1.397 \\
Unified OTF TF Gen & 0.697 & 0.089 \\
Twiddle Factor Seed Memory & 0.046 & 0.022 \\
MSE & 0.787 & 0.298 \\
PRNG & 0.069 & 0.028 \\
Local Scratchpad & 0.658 & 0.323 \\ \hline
\textbf{RSC} & \textbf{12.973} & \textbf{2.156} \\ \hline
2× RSC & 25.946 & 4.313 \\
Global Scratchpad & 2.632 & 1.290 \\
Top CTRL, DMA, Etc. & 0.060 & 0.051 \\ \hline
\textbf{Total} & \textbf{28.638} & \textbf{5.654} \\ \hline
\vspace{-0.35in}
\end{tabular}
\end{table}

We synthesize \ArchName's functional units, scratchpads, and interconnections using a 28nm process design kit, with a focus on minimizing area and power consumption while maintaining a 600 MHz clock frequency. All functional unit datapaths, including the RFE and MSE, support both 44-bit integer and 55-bit floating-point operations. The global scratchpad, designed as a double-buffered, single-port, multi-bank 256-bit width SRAM with a total capacity of 880 KB, stores messages received from the host and processed ciphertexts sent to the server. Assuming LPDDR5, which is commonly used in client-side environments with a bandwidth of 68.4 GB/s, this architecture effectively manages message fetching and ciphertext transmission. Additionally, each \CoreName incorporates a local scratchpad, implemented as a single-port, multi-bank 256-bit width SRAM. For PNL implementation, the shuffling unit requires FIFOs whose capacity doubles with each stage; these are also implemented as double-buffered SRAMs to minimize area overhead. Table \ref{table:area_power_breakdown} summarizes the synthesis results of \ArchName with two \CoreName instances. The overall design occupies 28.638 mm$^2$ and consumes up to 5.654 W of power. Based on the scaling methodology proposed in ~\cite{sarangi2021deepscaletool}, scaling to a 7nm process would reduce the area to approximately 0.9 mm$^2$ and the power consumption to 2.1 W, making \ArchName highly feasible for client-side applications.

\subsection{Evaluation Setup}
To evaluate the performance of \ArchName, we employed a two-step approach: a cycle-level simulator was developed to measure latency, and Design Compiler was used to determine area and power characteristics. For this evaluation, we selected parameters used in CKKS, including a polynomial degree of $2^{16}$. The bitwidth of the prime numbers was set to 36-bit, following the double scale technique~\cite{agrawal2023high}, and the number of levels was doubled from the standard 12 to 24. Additionally, we assumed that messages from the client are encrypted to a 24-level ciphertext, while ciphertexts from the server arrive at a 2-level state to facilitate efficient computation on the server and minimize computational overhead on the client.

\begin{figure}[t]
    \centering
    \includegraphics[width=3.4in]{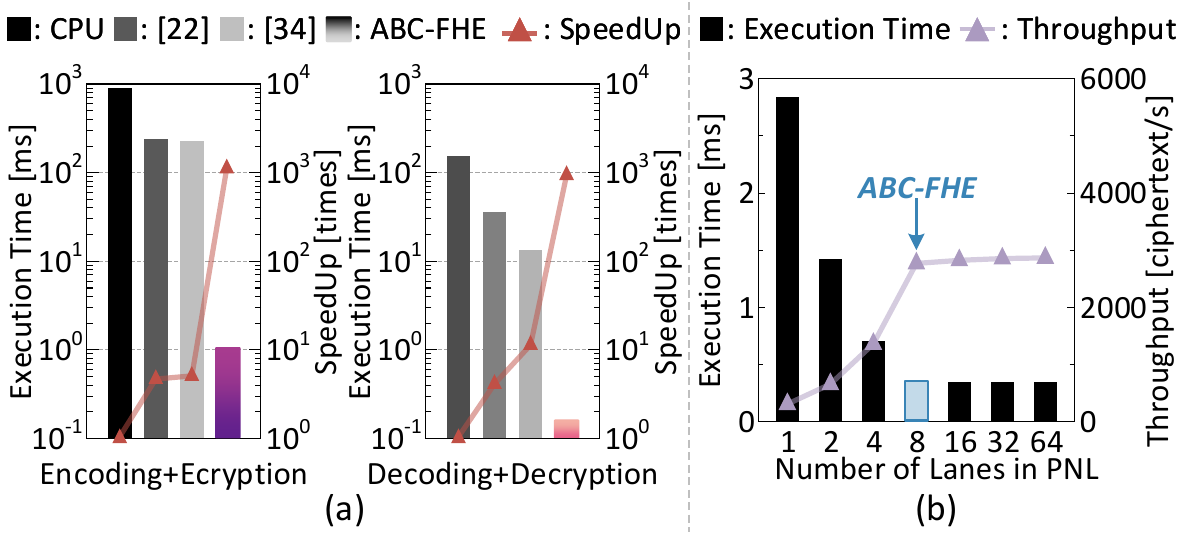}
    \vspace{-0.10in}
    \caption{Performance of \ArchName (a) Execution time and speed-up for encoding/encryption and decoding/decryption (b) Effect of number of pipelined NTT lanes on execution time and throughput.}
    \label{fig:performance}
    \vspace{-0.25in}
\end{figure}
\subsection{Results and Discussions}
\textbf{Latency and Throughput.}
Figure \ref{fig:performance} (a) highlights the significant performance improvements of \ArchName in terms of execution time and speed-up compared to prior works. Since previous designs do not support bootstrappable parameters, their reported latency was scaled by the proportion of operations for fair comparison. Additionally, to account for platform differences, FPGA and ASIC results were normalized to match \ArchName's 600 MHz frequency. \ArchName demonstrated a remarkable latency reduction: for encoding and encryption, it achieved a 1112$\times$ reduction compared to a PC-grade Intel Core i7-12700 using Lattigo~\cite{lattigo} and a 214$\times$ reduction compared to SOTA ASIC and FPGA implementations. For decoding and decryption, \ArchName reduced latency by 963$\times$ compared to the CPU and 82$\times$ compared to state-of-the-art ASIC and FPGA implementations. These results underscore \ArchName's superior efficiency, particularly in handling encoding and encryption tasks with large levels through streaming operations.

Figure \ref{fig:performance} (b) shows the impact of varying the number of PNLs on execution time and throughput. Under LPDDR5 specs typically used in client-side applications, the memory bottleneck was observed to cap performance at a maximum of 8 lanes, which \ArchName utilizes to maximize acceleration. By leveraging this optimal configuration, \ArchName maximized the number of ciphertexts processed per second, achieving peak performance in client-side FHE operations.

\begin{figure}[t]
    \centering
    \includegraphics[width=3.4in]{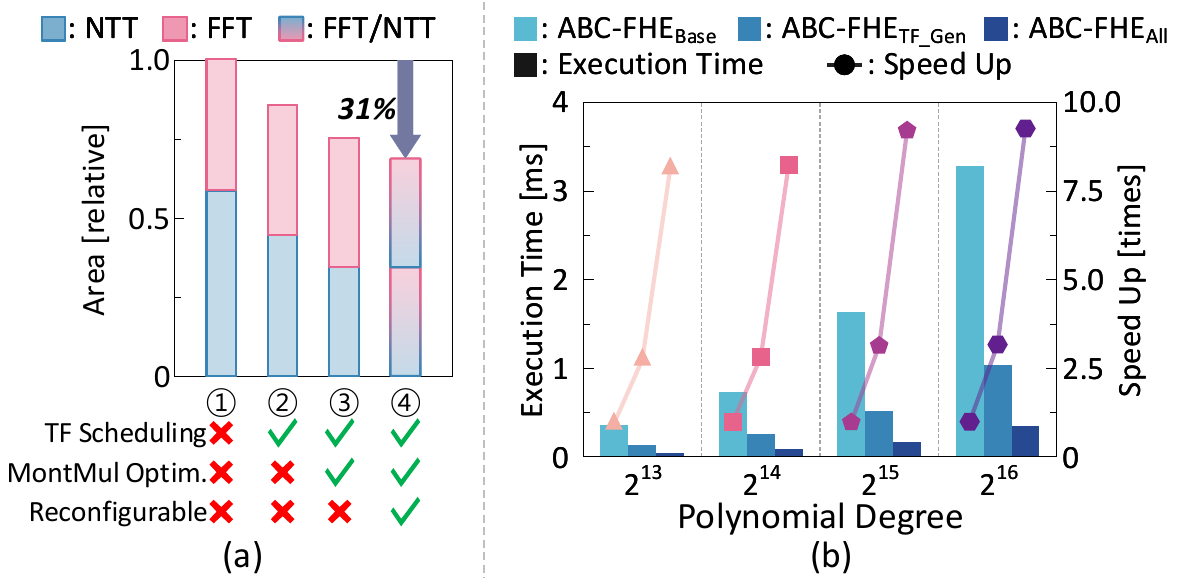}
    \vspace{-0.10in}
    \caption{(a) Area reduction as area optimization applied (b) Execution time improvement as memory optimization applied}
    \label{fig:ablation_study}
    \vspace{-0.25in}
\end{figure}

\textbf{Area Optimization in RFE.}
Figure \ref{fig:ablation_study} (a) demonstrates the effectiveness of the proposed optimizations in reducing the hardware area of the RFE. All designs are evaluated using a P=8 MDC-based pipelined NTT architecture, comparing the area required to generate one FFT result and four NTT results. For fairness, the comparison assumes two RFEs for cases requiring separate hardware for NTT and FFT computations, with other configurations adjusted accordingly. 

The baseline design adopts the commonly used radix-2 pipelined NTT architecture with distinct hardware implementations for NTT and FFT. Optimization techniques include twiddle factor scheduling to reduce the number of multipliers and Montgomery multiplier optimization to minimize the area of computational units. The final configuration represents a fully reconfigurable RFE capable of supporting both NTT and FFT operations. Combined, these optimizations achieved a 31\% reduction in total area, underscoring the efficiency of the proposed approach.

\textbf{On-chip Memory Optimization.}
To minimize EMA, \ArchName incorporates a unified OTF TF Gen and a PRNG to generate critical data such as twiddle factors, masks, errors, and keys directly on-chip. Figure \ref{fig:ablation_study} (b) evaluates the performance gain achieved by this approach across various polynomial degrees. Three configurations were analyzed: ABC-FHE$_{\text{Base}}$, which fetches all necessary data from external memory; ABC-FHE$_{\text{TF\_Gen}}$, which generates only twiddle factors on-chip; and ABC-FHE$_{\text{All}}$, which uses both unified OTF TF Gen and PRNG to generate all data on-chip. Compared to external memory reliance, ABC-FHE$_{\text{All}}$ achieved a latency reduction of approximately $8.2$–$9.3\times$. Given that the combined area of the unified OTF TF Gen and PRNG constitutes only 6\% of the total chip area, this design demonstrates significant performance improvements with minimal area overhead.



\section{Conclusion}
\label{section6}

In this work, we proposed \ArchName, a resource-efficient client-side accelerator designed to support bootstrappable parameters. Prior works fail to address bootstrappable parameters or rely on non-streaming architectures, leading to memory bottlenecks and latency overhead when throughput increases. To overcome these issues, \ArchName employs a streaming architecture for consistent throughput and generates parameters like random values and twiddle factors on-chip, minimizing external memory access. This design achieves significant speedups: 1112$\times$ and 214$\times$ faster encoding/encryption and 963$\times$ and 82$\times$ faster decoding/decryption compared to a CPU and the state-of-the-art accelerator, respectively.

\section*{Acknowledgment}
This work was supported by the MSIT (Ministry of Science and ICT), Korea, through the IITP (Institute for Information \& Communications Technology Planning \& Evaluation) under the ITRC program (IITP-2025-RS-2020-II201847), the PIM Technology Development project (2022-0-01037), and the Graduate School of AI Semiconductor program (IITP-2025-RS-2023-00256472).

\newpage
\bibliographystyle{IEEEtran}
\footnotesize


\end{document}